\documentstyle[12pt,fleqn,epsf]{article}
\def\ap#1#2#3{     {\it Ann. Phys. (NY) }{\bf{#1},} (19#3) #2 }
\def\jmp#1#2#3{   {\it J. Math. Phys. } {\bf{#1},}  (19#3) #2 }

\def\amjp#1#2#3{  {\it Am. J. Phys.} {\bf{#1},} (19#3) #2 }

\def\fderp#1#2#3{  {\it Fortsch. der Phys.} {  \bf{#1},} (#3) #2 }

\def\pla#1#2#3{    {\it Phys. Lett. }{ A \bf{#1},} (#3) #2 }
\def\prd#1#2#3{    {\it Phys. Rev. }{ D \bf{#1},} (19#3) #2 }
\def\pra#1#2#3{    {\it Phys. Rev. }{ A \bf{#1},} (#3) #2 }

\def\reppp#1#2#3{  {\it Rep. Prog. Phys.} {\bf{#1},} (19#3) #2  }
\def\prl#1#2#3{    {\it Phys. Rev. Lett. }{\bf{#1},} (19#3) #2 }

\def\mpla#1#2#3{    {\it Mod. Phys. Lett. }{A \bf{#1},} (#3) #2 }

\def\praj#1#2#3{    {\it Pramana. J. Phys. }{\bf{#1},} (#3) #2 }

\def\eq#1{{eq.~(\ref{#1})}}

\newcommand{\bea}{\begin{eqnarray}}
\newcommand{\beq}{\begin{equation}}
\newcommand{\eea}{\end{eqnarray}}
\newcommand{\eeq}{\end{equation}}

\begin{document}
\title{\bf Spectrum of One-Dimensional Multiple Well  Oscillators }
{\small
\author{H. A. Alhendi$^1$\ and \ E. I. Lashin$^{1,2}$\\
$^1$ Department of physics and Astronomy, College of Science,\\ King Saud University, Riyadh,
Saudi Arabia \\
$^2$ Department of Physics, Faculty of Science, \\Ain Shams University, Cairo, Egypt}
}
\maketitle
\begin{abstract}
We apply power series expansion to symmetric  multi-well oscillators bounded by two
infinite walls. The spectrum and expectation values  obtained are compared with available exact
and approximate values for the unbounded ones. It is shown that the method is capable of
producing to a high accuracy the eigenvalues,
eigenfunctions, and the expectation values $\langle x^{2k}\rangle$ of the corresponding unbounded
ones as the separation between the two infinite walls becomes large.
\end{abstract}
\section{Introduction}
Potential energy functions with degenerate classical minima
separated by potential barriers are frequently encountered in many
areas of physics, such as molecular physics, nonlinear physics, field theory, and
cosmology. A widely investigated model, by variant methods
(\cite{froman-dw}--\cite{ours-dw}), is a one-dimensional symmetric double-well potential, in
which the degeneracy of the low-lying doublets is broken by
quantum tunneling effects \cite{landau}.

Recently, calculation of  energy splitting of a one-dimensional
symmetric  multi-well potential energy functions has received a
considerable interest (\cite{adh-mw}--\cite{gupta-mw}). This interest stems from the fact
that the topology of the multi-well potential is
completely different from the double-well case. Each well, except the two outer ones,
is surrounded by two barriers. In the case of odd number of wells,
the positions of the minima of the potential are connected by the spatial
inversion symmetry of the potential that leaves the position of the minimum at
the origin invariant, for even number of wells ,however, these
minima are connected by the inversion symmetry. These topological
properties are relevant for calculations based on path integral
methods using instanton solutions\cite{coleman-dw,raj,klen}.

In an earlier paper \cite{ours-dw}, we have applied the traditional
power series expansion method with boundary conditions in the
finite interval to one-dimensional anharmonic potential functions
having single and double-well. We have shown that the method is
capable, even for a reasonable size  of the interval, of producing
to a high accuracy the eigenvalues and eigenfunctions of the
corresponding unbounded ones. In particular for the deep double-well
case, with potential function $(V(x)=-\mu^2 x^2 + x^4)$, we have
shown that the splitting of the energy levels of the ground
and first excited states  becomes apparent only
after $26$ significant digits for the mass parameter $\mu^2=25$, and for $\mu^2=35$ the
splitting shows up after $42$ significant digits, while for $\mu^2=50$
the splitting starts after $72$ significant digits.

In the present paper we extend our previous calculation to
multi-well potentials and compare with available exact and
approximate values recently obtained by other methods.

The rest of the paper is organized as follows. In sec.~2 we apply
the method of  power series expansion to anharmonic multi-well
potentials, and present our calculation of the eigenvalues, eigenfunctions, and
expectation values of even powers of the x-coordinate and  compare  with
that recently obtained  by the improved Hill determinant method \cite{adh-mw,varma-mw},
and with the exact
values calculated by supersymmetric quantum mechanics \cite{adh-mw}.
It is shown that the method
yields extremely very close results to the exact ones. Finally  in
sec.~3 we give our conclusion.

\section{Calculation of energies and expectation values}
In order to calculate the eigenvalues and eigenfunctions of the
anharmonic multi-well potential bounded by infinitely high
potential walls located at $x=\pm L$, one needs to solve the eigenvalue equation
(in units $\hbar=1\, ,\, 2\,m=1$) :
\beq
\left[\frac{d^2}{d\,x^2}+E-V(x)\right]\Psi(x)=0 ,
 \label{seq}
 \eeq
with the boundary conditions $\Psi(\pm L)=0$. In the present work we consider
potentials $V(x)$ of the form \cite{chmon-mw,varma-mw,gupta-mw}:
\beq
V(x)=\sum_{k=1}^{N} b_{2k} x^{2k},\hspace{2cm}
\label{pot}
\eeq
Here the coefficients ($b_{2m}$) are real with $b_{2N}$ being
positive, and $N=3,4,5.$

Making use of the power series expansion
\beq
\Psi = \sum_{n=0}^{\infty} a_n x^n
\label{series}
\eeq
in \eq{seq} we obtain the following recurrence formula for the expansion coefficients:
\bea
a_n = \frac{-E\, a_{n-2} + b_2\, a_{n-4} + b_4\, a_{n-6}\,
+ b_6\, a_{n-8}+ b_8\, a_{n-10}+b_{10}\, a_{n-12}}{n\,(n-1)}, & & n \neq 0,1 \nonumber \\
a_n  =  0, & & n < 0 \nonumber\\
\label{rec}
\eea
The symmetry of \eq{seq} divides the solutions
into two types even and odd. The even solutions can be obtained by
imposing (ignoring normalization) $a_{0}~=~1, a_{1}~=~0 $, while the odd ones by imposing
$a_{0}~=~0, a_{1}~=~1$. The energy eigenvalues $(E)$ are obtained from
the condition $\Psi(L)=0$ in both cases. Since
we are dealing with potentials admitting power series expansion for
$|x|<L$, the power series solutions of $\Psi(x)$ are,
according to a well known theorem in differential equations, convergent \cite{codd}.
Here we present our calculation of the energy levels, wave functions and
expectation values for
the potentials given in \eq{pot}, using
eqs.(\ref{series}),(\ref{rec})
and the boundary condition at $x=L$.

Table~\ref{tabev1} shows the first four energy eigenvalues of three-, four-, and five-well
oscillators  for the bounded  potential-wells
as compared with the obtained values reported in \cite{chmon-mw,varma-mw}
for the unbounded case. The obtained values are extremely very close to the exact ones.
{\small
\begin{table}[htbp]
\begin{center}
\begin{tabular}{ccccc}
\hline
\hline
          &          &           &             &Results of \\
          &  No. of  &           &             &supersymmetric \\
Potential & wells    & $L$           & Our results & quantum mechanics\\
\hline
$A(x)$   & 3  & $4$ &$\underline{-2.000\ 000\ 000\ 000}\ 000\ 000\ $ &$-2$\\
         &    &  &$\underline{-1.772\ 726\ 698\ 991}\ 350\ 330\ $ &    \\
         &    & & $\underline{ 2.078\ 279\ 891\ 768}\ 595\ 361\ $ &     \\
         &    &  &$\underline{ 5.604\ 028\ 342\ 382}\ 013\ 654\ $ &\\
\\
$B(x)$   & 3  & $4$ & $\underline{-9.001\ 720\ 238\ 527}\ 719\ 715\ $ &  \\
         &    & &$\underline{-9.000\ 000\ 000\ 000}\ 000\ 000\ $ & $-9$ \\
         &    & &$\underline{ 0.639\ 394\ 262\ 865}\ 333\ 280\ $ &  \\
         &    & &$\underline{ 1.936\ 629\ 224\ 926}\ 380\ 607\ $ &\\
\\
$C(x)$& $3$   & $3$ & $\underline{0.375\ 000\ 000\ 000}\ 000\ 000\ $ & $0.375$\\
      &       & &$\underline{2.357\ 398\ 881\ 839}\ 175\ 696\ $ &  \\
      &       & &$\underline{6.988\ 755\ 014\ 467}\ 723\ 300\ $ & \\
      &       & &$\underline{13.880\ 516\ 236\ 71}   388\ 190\ $ & \\
\\
$D(x)$& $3$  & $3$ & $\underline{-0.195\ 122\ 059\ 734}\ 627\ 597\ $ & \\
      &      & &$\underline{ 1.125\ 000\ 000\ 000}\ 000\ 000\ $ &$1.125$ \\
      &      & &$\underline{ 5.646\ 143\ 524\ 135}\ 629\ 302\ $ &\\
      &      & &$\underline{12.384\ 746\ 745\ 98}\  872\ 574\ $  &\\
\\
$F(x)$& $4$ & $4$ & $\underline{-0.223\ 991\ 055\ 384}\ 171\ 854\ $ &\\
      &     &  &$\underline{ 0.083\ 481\ 557\ 863}\ 966\ 793\ $ &\\
      &     &  &$\underline{ 1.526\ 487\ 708\ 073}\ 844\ 797\ $ &\\
      &     &  &$\underline{ 3.971\ 174\ 256\ 474}\ 939\ 459\ $ &\\
\\
$G(x)$& $4$ & $3$ & $\underline{-0.096\ 291\ 946\ 230}\ 649\ 098\ $ &\\
      &     &  &$\underline{ 0.672\ 993\ 242\ 745}\ 446\ 704\ $ & \\
      &     &  &$\underline{ 3.111\ 022\ 328\ 724}\ 771\ 653\ $ &  \\
      &     &  &$\underline{ 7.038\ 082\ 659\ 880}\ 398\ 654\ $ &  \\
\\
$H(x)$& $5$ & $4$& $\underline{ 0.807\ 741\ 647\ 209}\ 432\ 443\ $ & \\
      &     &  &$\underline{ 3.277\ 946\ 311\ 571}\ 061\ 982\ $ & \\
      &     &  &$\underline{ 7.667\ 480\ 496\ 116}\ 480\ 534\ $ & \\
      &     &  &$\underline{13.578\ 984\ 131\ 990}\ 285\ 801\ $ &\\
\hline
\hline
\end{tabular}
\end{center}
\caption{{\small The first four energy eigenvalues of three-, four-, and
five-well oscillators: $A(x)=x^2-4 x^4 + x^6,\ B(x)=4 x^2 - 6 x^4 +x^6,\
C(x)={105\over 64} x^2 - {43\over 8} x^4 + x^6 - x^8+ x^{10},\
D(x)={169\over 64} x^2 - {59\over 8} x^4 + x^6 - x^8+ x^{10},\
F(x)=-x^2+2 x^4 - 0.9 x^6 + 0.1 x^8,\ G(x)=-x^2+3 x^4 - 2 x^6 + 0.1 x^{10},\
H(x)=2 x^2 - 7.5 x^4 + 5.5 x^6 -0.877 x^8 + 0.04 x^{10}$ as obtained in the
present work for the bounded cases ($2L$ is the width of the well).
The underlined numbers correspond to the unbounded ones \cite{varma-mw}. }}
\label{tabev1}
\end{table}
}
{\small
\begin{table}[htbp]
\begin{center}
\begin{tabular}{ccccc}
\hline
\hline
Potential      &  Expectation        &             & results & \\
and eigenstate &  value              &Our results  &  in \cite{chmon-mw}& Exact\\
\hline
$A(x), \Psi_0(x)$ & $\langle x^2 \rangle$ &$1.7042723043$ &$1.704$ &$1.7043$ \\
                  &$\langle x^4 \rangle$  &$3.9085446087$ &$3.908$ & $3.9085$ \\
                  &$\langle x^6 \rangle$& $10.373497673$  &$10.349$ &  $10.3735$    \\
                  &$\langle x^8 \rangle$ &$30.518356869$  &$30.212$ & $30.5184$  \\
                 &$\langle x^{10}\rangle$&$97.343955598$  &$93.210$ & $97.3440$  \\
\\
$B(x), \Psi_1(x)$ & $\langle x^2 \rangle$ &$3.1795525642$  &$3.179$ &$3.1796$ \\
                  &$\langle x^4 \rangle$ &$11.0386576927$  &$11.035$ &$11.0387$  \\
                  &$\langle x^6 \rangle$ &$41.0648544887$  &$40.985$ & $41.0649$     \\
                  &$\langle x^8 \rangle$&$161.8298653908$  &$159.909$ &  $161.8299$ \\
               &$\langle x^{10} \rangle$&$670.2814413720$ &$625.164$ & $670.82814$  \\
\\
$C(x), \Psi_0(x)$ & $\langle x^2 \rangle$ &$0.45832470069$  & $0.458$&$0.4583$ \\
                  &$\langle x^4 \rangle$  &$0.43854420934$  &$0.439$ & $0.4385$ \\
                  &$\langle x^6 \rangle$  &$0.54740034191$  &$0.547$ & $0.5474$     \\
                  &$\langle x^8 \rangle$  &$0.79673314349$  &$0.796$ & $0.7967$  \\
                 &$\langle x^{10} \rangle$&$1.28945196690$  &$1.288$ &$1.2895$   \\
\\
$D(x), \Psi_1(x)$ & $\langle x^2 \rangle$ &$0.956841751448$  &$0.957$ &$0.9568$ \\
                  &$\langle x^4 \rangle$  &$1.194350514116$  &$1.194$ &$1.1944$  \\
                  &$\langle x^6 \rangle$  &$1.738359600265$  &$1.738$ & $1.7384$     \\
                  &$\langle x^8 \rangle$  &$2.8134027359576611$  &$2.812$ & $2.8134$ \\
               &$\langle x^{10} \rangle$  &$4.935043317278$  &$4.930$ & $4.9350$  \\
\hline
\hline
\end{tabular}
\end{center}
\caption{{\small Comparison of the expectation values $\langle x^{2m}\rangle$ of
the supersymmetric potential $A(x), B(x), C(x),\mbox{and} D(x)$ obtained in
the present work  for the bounded cases and that of ref.~\cite{chmon-mw} for the unbounded ones
along with  the exact values obtained using supersymmetric quantum mechanics
\cite{adh-mw,chmon-mw} }}
\label{tabev2}
\end{table}
}
Table~\ref{tabev2} contains a comparison of the expectation values $\langle x^{2m}\rangle,
(m=1\cdots 5)$ with the reported ones and  the exact values
obtained from the corresponding supersymmetric  wave functions.
For completeness we also present in Fig.~\ref{figmw}~$(A,C)$ the ground state wave functions for the
potentials $A(x)$ and $C(x)$ respectively, and~$(B,D)$ are  for the first excited state wave
functions for the potentials $B(x)$  and $D(x)$ respectively.
These coincide  with the corresponding exact wave functions (ignoring normalization) obtained using supersymmetric
quantum mechanics \cite{adh-mw}:
\begin{figure}[htbp]
\epsfxsize=15cm
\centerline{\epsfbox{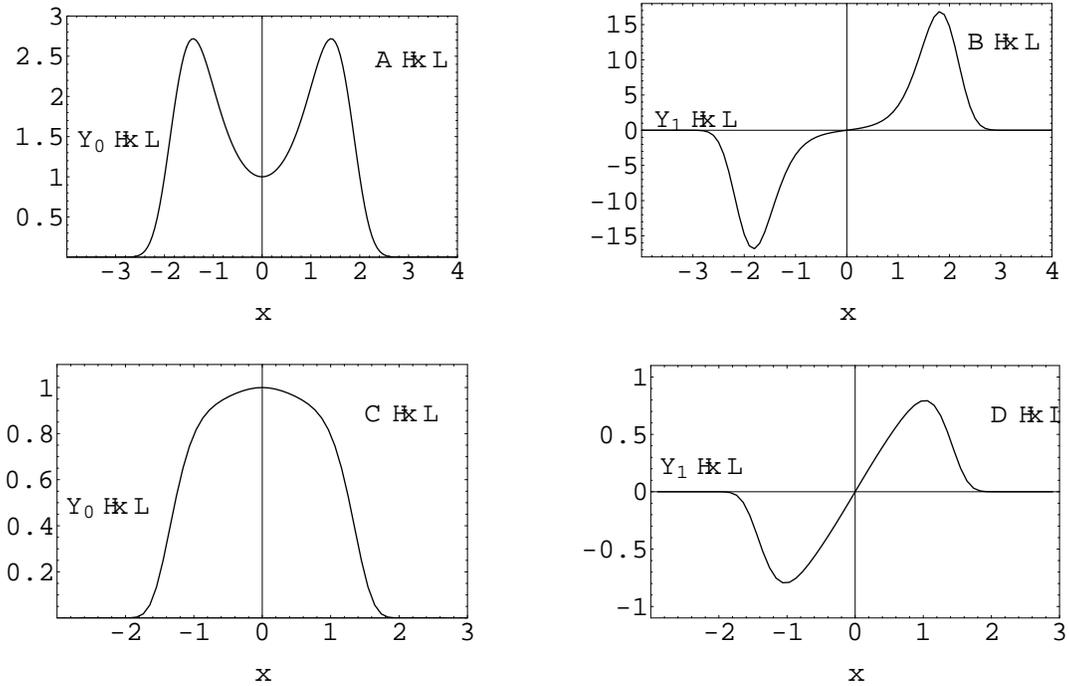}}
\caption{These graphs show  the ground state wave functions for the potentials
$A(x)$ and $C(x)$, also showing the first excited state wave functions corresponding
to the potentials $B(x)$ and $D(x)$ for the bounded case}
\label{figmw}
\end{figure}
$$
\begin{array}{llll}
\Psi_0(x)&=&\exp{(x^2-x^4/4)} & \mbox{for}\ \ A(x)\\
\Psi_0(x)&=&\exp{(-3 x^2/16 + x^2/8 - x^6/6)} & \mbox{for}\ \ C(x)\\
\Psi_1(x)&=&x\ \exp{(3 x^2/2 - x^4/4)} & \mbox{for}\ \ B(x)\\
\Psi_1(x)&=& x\ \exp{(-3 x^2/16 + x^4/8 - x^6/6)} & \mbox{for}\ \ D(x)\\
\end{array}
$$
These calculations show that even for a reasonable size of the interval the present method,
beside being simple and direct is  highly effective.
\section{Conclusion}
In this paper we have applied the method of power series expansion to a
variant of one-dimensional multi-well potential functions
bounded by two infinite walls. We have compared our calculation
of the low-lying energy levels, wave functions, and expectation values
with that obtained using improved Hill determinant and
supersymmetric quantum mechanics. For  all these potentials  we
have obtained results that are extremely very close to the exact ones.
This is because the power series expansion in the finite range is
convergent for potentials admitting power series expansion.
It is easy to extend the present method to other multi-well
polynomial potential functions with higher degrees and more degenerate classical
minima.

\begin{flushleft}
{\bf Acknowledgement}
\end{flushleft}
This work was supported by research center at college of science,
King Saud University under project number Phys$/1423/02$.
\newpage
\clearpage
\renewcommand{\baselinestretch}{1}

\end{document}